\documentstyle{aipproc}
\def\bd{B_d^0}
\def\bdbar{{\overline{B_d^0}}}
\def\bs{B_s^0}
\def\bsbar{{\overline{B_s^0}}}
\def\sss{\scriptscriptstyle}
\def\ks{K_{\sss S}}
\newcommand{\beq}{\begin{equation}}
\newcommand{\eeq}{\end{equation}}
\def\ijmp#1#2#3{{\it Int.\ J.\ Mod.\ Phys.} {\bf A#1}, #3 (19#2)}

\def\npb#1#2#3{{\it Nucl.\ Phys.} {\bf B#1}, #3 (19#2)}
\def\plb#1#2#3{{\it Phys.\ Lett.} {\bf B#1}, #3 (19#2)}
\def\prd#1#2#3{{\it Phys.\ Rev.} {\bf D#1}, #3 (19#2)}

\def\prl#1#2#3{{\it Phys.\ Rev.\ Lett.} {\bf #1}, #3 (19#2)}
\def\zpc#1#2#3{{\it Zeit.\ Phys.} {\bf C#1}, #3 (19#2)}
\def \stone{{\it B Decays}, edited by S. Stone 
(World Scientific, Singapore, 1994)}
\newread\epsffilein 
\newif\ifepsffileok 
\newif\ifepsfbbfound 
\newif\ifepsfverbose 
\newdimen\epsfxsize 
\newdimen\epsfysize 
\newdimen\epsftsize 
\newdimen\epsfrsize 
\newdimen\epsftmp  
\newdimen\pspoints  
\def\epsfbox#1{\global\def\epsfllx{72}\global\def\epsflly{72}%
 \global\def\epsfurx{540}\global\def\epsfury{720}%
 \def\lbracket{[}\def\testit{#1}\ifx\testit\lbracket
 \let\next=\epsfgetlitbb\else\let\next=\epsfnormal\fi\next{#1}}%
\def\epsfgetlitbb#1#2 #3 #4 #5]#6{\epsfgrab #2 #3 #4 #5 .\\%
 \epsfsetgraph{#6}}%
\def\epsfnormal#1{\epsfgetbb{#1}\epsfsetgraph{#1}}%
\def\epsfgetbb#1{%
\openin\epsffilein=#1
\ifeof\epsffilein\errmessage{I couldn't open #1, will ignore it}\else
 {\epsffileoktrue \chardef\other=12
 \def\do##1{\catcode`##1=\other}\dospecials \catcode`\ =10
 \loop
  \read\epsffilein to \epsffileline
  \ifeof\epsffilein\epsffileokfalse\else
   \expandafter\epsfaux\epsffileline:. \\%
  \fi
 \ifepsffileok\repeat
 \ifepsfbbfound\else
 \ifepsfverbose\message{No bounding box comment in #1; using defaults}\fi\fi
 }\closein\epsffilein\fi}%
\def\epsfclipstring{}
\def\epsfsetgraph#1{%
 \epsfrsize=\epsfury\pspoints
 \advance\epsfrsize by-\epsflly\pspoints
 \epsftsize=\epsfurx\pspoints
 \advance\epsftsize by-\epsfllx\pspoints
 \epsfxsize\epsfsize\epsftsize\epsfrsize
 \ifnum\epsfxsize=0 \ifnum\epsfysize=0
  \epsfxsize=\epsftsize \epsfysize=\epsfrsize
  \epsfrsize=0pt
  \else\epsftmp=\epsftsize \divide\epsftmp\epsfrsize
  \epsfxsize=\epsfysize \multiply\epsfxsize\epsftmp
  \multiply\epsftmp\epsfrsize \advance\epsftsize-\epsftmp
  \epsftmp=\epsfysize
  \loop \advance\epsftsize\epsftsize \divide\epsftmp 2
  \ifnum\epsftmp>0
   \ifnum\epsftsize<\epsfrsize\else
    \advance\epsftsize-\epsfrsize \advance\epsfxsize\epsftmp \fi
  \repeat
  \epsfrsize=0pt
  \fi
 \else \ifnum\epsfysize=0
  \epsftmp=\epsfrsize \divide\epsftmp\epsftsize
  \epsfysize=\epsfxsize \multiply\epsfysize\epsftmp
  \multiply\epsftmp\epsftsize \advance\epsfrsize-\epsftmp
  \epsftmp=\epsfxsize
  \loop \advance\epsfrsize\epsfrsize \divide\epsftmp 2
  \ifnum\epsftmp>0
  \ifnum\epsfrsize<\epsftsize\else
   \advance\epsfrsize-\epsftsize \advance\epsfysize\epsftmp \fi
  \repeat
  \epsfrsize=0pt
 \else
  \epsfrsize=\epsfysize
 \fi
 \fi
 \ifepsfverbose\message{#1: width=\the\epsfxsize, height=\the\epsfysize}\fi
 \epsftmp=10\epsfxsize \divide\epsftmp\pspoints
 \vbox to\epsfysize{\vfil\hbox to\epsfxsize{%
  \ifnum\epsfrsize=0\relax
  \includegraphics{#1}%
  \else
  \epsfrsize=10\epsfysize \divide\epsfrsize\pspoints
  \includegraphics{#1}%
  \fi
  \hfil}}%
\global\epsfxsize=0pt\global\epsfysize=0pt}%
 {\catcode`\%=12 \global\let\epsfpercent=
\long\def\epsfaux#1#2:#3\\{\ifx#1\epsfpercent
 \def\testit{#2}\ifx\testit\epsfbblit
  \epsfgrab #3 . . . \\%
  \epsffileokfalse
  \global\epsfbbfoundtrue
 \fi\else\ifx#1\par\else\epsffileokfalse\fi\fi}%
\def\epsfempty{}%
\def\epsfgrab #1 #2 #3 #4 #5\\{%
\global\def\epsfllx{#1}\ifx\epsfllx\epsfempty
  \epsfgrab #2 #3 #4 #5 .\\\else
 \global\def\epsflly{#2}%
 \global\def\epsfurx{#3}\global\def\epsfury{#4}\fi}%
\def\epsfsize#1#2{\epsfxsize}
\begin{document}
\begin{flushright}
UdeM-GPP-TH-97-44 \\
hep-ph/970xxxx \\
August 1997 \\
\end{flushright}

\title{New Physics and the \\ Unitarity Triangle\thanks{Invited talk given
at the Symposium {\it Twenty Beautiful Years of Bottom Physics}, Chicago,
IL, USA, June 29 -- July 2, 1997.}}

\author{David London}
\address{Laboratoire de physique nucl\'eaire, Universit\'e de Montr\'eal,
\\ 
C.P. 6128, succ.\ centre-ville, Montr\'eal, QC, Canada}

\maketitle

\begin{abstract}
After reviewing the present experimental constraints on the unitarity
triangle, I discuss the various ways in which new physics can manifest
itself in measurements of the parameters of the unitarity triangle. Apart
from one exception, which I describe, new physics enters principally
through new contributions to $B^0$-${\overline{B^0}}$ mixing. Different
models of new physics can be partially distinguished by looking at their
effects on rare, flavour-changing $B$ penguin decays.
\end{abstract}

At this conference, we have heard a number of talks discussing the
prospects of various experiments for measuring CP asymmetries in $B$
decays, {\it i.e.} the angles of the unitarity triangle. Ultimately, the
hope is that we will find an inconsistency with the standard model (SM),
which will give us some clue regarding the new physics which most of us
believe must lie beyond the SM. In discussing new physics and the unitarity
triangle (UT), there are basically two questions which have to be
addressed: 
\begin{enumerate}

\item What are the signals of new physics?

\item If such signals are seen, how can we identify the new physics?

\end{enumerate}

The first step in answering these questions is to review our current
knowledge of the UT. There are a number of measurements which constrain the
UT: $|V_{cb}|$, $|V_{ub}/V_{cb}|$, $B_d$ and $B_s$ mixing, and $|\epsilon|$
in the kaon system. However, the problem is that there are important
theoretical uncertainties in translating the experimental numbers into
information about the UT. Combining all theoretical and experimental errors
in quadrature, our present knowledge of the UT can be summed up in Fig.~1
\cite{London:AliLon}. As is evident from this figure, we really know rather
little about the UT at present, due mainly to theoretical uncertainties.


\begin{figure}
\vskip -1.0truein
\centerline{\epsfxsize 3.5 truein \epsfbox {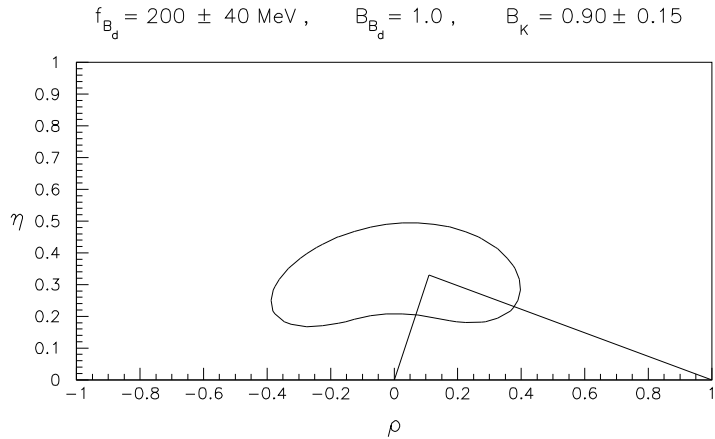}}
\vskip -1.4truein
\caption{Allowed region in $\rho$-$\eta$ space, from a simultaneous fit to
both the experimental data (as given in Ref.~\protect\cite{London:AliLon})
and theoretical quantities (listed above). The theoretical errors are
treated as Gaussian. The solid line represents the 95\% c.l.\ region. The
triangle shows the best fit.} 
\end{figure}

On the other hand, the angles $\alpha$, $\beta$ and $\gamma$ can be
extracted with essentially no theoretical uncertainty from CP-violating
asymmetries in $B$ decays \cite{London:CPReview}. Due to
$B^0$-${\overline{B^0}}$ mixing, any neutral $B$ decay to a final state $f$
to which both $B^0$ and ${\overline{B^0}}$ can decay can exhibit CP
violation, due to the interference between the amplitudes $B^0 \to f$ and
${\overline{B^0}} \to f$. There are 4 distinct classes of CP asymmetries,
involving the decays of $\bd$ or $\bs$ mesons, and the quark-level
processes $b\to c$ or $b\to u$. For example, the asymmetries in $\bd(t) \to
\pi^+\pi^-$ and $\bd(t) \to \Psi\ks$ probe $\sin 2\alpha$ and $\sin
2\beta$, respectively, while $\sin^2 \gamma$ can be extracted from $\bs(t)
\to D_s^\pm K^\mp$. (The decay $B^\pm \to D K^\pm$ can also be used to
obtain $\sin^2 \gamma$.) The fourth CP asymmetry ({\it e.g.} $\bs(t) \to
\Psi\phi$) is expected to be zero, to a good approximation, within the SM.
This is therefore a good place to look for new physics. The measurements of
the three nonzero CP angles would allow us to reconstruct the UT with
little theoretical error. Our present experimental knowledge constrains the
angles to lie within the ranges $-1.0 \le \sin 2\alpha \le 1.0$, $0.26 \le
\sin 2\beta \le 0.88$, and $0.22 \le \sin^2 \gamma \le 1.0$. 

There are thus three distinct ways in which new physics can manifest itself
in measurements of these CP asymmetries \cite{London:GL}:
\begin{enumerate}

\item $\alpha+\beta+\gamma \ne \pi$.

\item $\alpha+\beta+\gamma=\pi$, but the values of $\alpha$, $\beta$ and
$\gamma$ found disagree with the SM predictions.

\item $\alpha+\beta+\gamma=\pi$, $\alpha$, $\beta$ and $\gamma$ are
consistent with the SM, but measurements of the angles are inconsistent
with measurements of the {\it sides} of the UT.

\end{enumerate}

Now, how can new physics affect these CP asymmetries? There are basically
only two ways, via either new contributions to $B$ decays ($b \to c,u$), or
to $B$-${\overline B}$ mixing. The first possibility can be virtually
eliminated -- apart from some very fine-tuned models, there are no models
of physics beyond the SM in which the new contributions are competitive
with the SM $W$-mediated decays. On the other hand, there are many models
of new physics in which there are new contributions to $B$-${\overline B}$
mixing, possibly with new phases \cite{London:DibLN}. Therefore {\it the
principal way in which new physics can affect the UT is via new
contributions to $B$ mixing.} (There is an exception to this, which I will
discuss below.) These new contributions will affect the experimental
determinations of $V_{td}$, $V_{ts}$, $\alpha$, $\beta$ and $\gamma$. 

In light of this, let us reconsider the three ways in which new physics can
be detected. The first is to measure the 3 CP angles, and find $\alpha +
\beta + \gamma \ne \pi$. In order for this to happen, there must be new
physics, {\it with new phases}, in $B_d$ or $B_s$ mixing. However, there is
a interesting twist here. Suppose there is new physics in $B$ mixing. If
$\beta$ is measured in $B_d(t) \to \Psi \ks$, then the phase extracted will
be $\beta + \phi^d_{\sss NP}$. And if $\alpha$ is obtained via $B_d(t) \to
\pi^+\pi^-$, then one gets $\alpha - \phi^d_{\sss NP}$. The key point here
is that the sum $\alpha+\beta$ is {\it insensitive} to new physics
\cite{London:NirSilv}. Turning to the third angle, if $\gamma$ is measured
in $B^\pm \to D K^\pm$, then it is extracted with no modification, since
neutral $B$'s are not involved. However, if $\gamma$ is obtained from
$B_s(t) \to D_s^\pm K^\mp$, then $\gamma + \phi^s_{\sss NP}$ will be
extracted. The upshot is: since $B$-factories such as BaBar and Belle do
not measure CP asymmetries in $\bs$ decays, they will {\it never} find
$\alpha + \beta + \gamma \ne 0$. (Once again, there is an exception, to be
discussed below.) However, hadron colliders may find $\alpha + \beta +
\gamma \ne 0$ if $\gamma$ is measured in $\bs$ decays. In fact, a
discrepancy in the value of $\gamma$ as extracted in these two ways would
be a clear signal for new physics in $\bs$-$\bsbar$ mixing.

The second way to detect new physics is if $\alpha+\beta+\gamma=\pi$, but
the values of $\alpha$, $\beta$ and $\gamma$ are in disagreement with the
SM predictions. This can happen if there are new contributions, {\it with
new phases}, to $B_d$ or $B_s$ mixing. Finally, the third way is if $\alpha
+ \beta + \gamma = \pi$ and $\alpha$, $\beta$ and $\gamma$ are consistent
with the SM, but are inconsistent with measurements of the sides of the UT.
In this case, we need new contributions to $B_d$ or $B_s$ mixing {\it with
the same phase} as in the SM.

Before examining which types of physics can contribute to $B$-${\overline
B}$ mixing, let me first discuss the exception I mentioned above. Most CP
asymmetries involve tree-level $B$ decays. However, there is another class
of decays which can also be used: penguin decays
\cite{London:LonPec,London:penguins}. Consider, for example, the decay $\bd
\to \phi \ks$, which is dominated by the quark-level ${\bar b} \to {\bar s}
s {\bar s}$ penguin decay. Since the final state is a CP eigenstate, both
$\bd$ and $\bdbar$ can decay to it, thus leading to a possible CP-violating
asymmetry. What does this CP asymmetry measure? The $b\to s$ penguin is
dominated by internal $t$-quarks, so that it is proportional to the product
of CKM matrix elements $V_{tb}^* V_{ts}$. Within the Wolfenstein
approximation, this is real, just like $V_{cb}^* V_{cs}$, which describes
the decay $\bd \to \Psi \ks$. In other words, the CP asymmetry in $B_d(t)
\to \phi \ks$ measures $\beta$, just like $B_d(t) \to \Psi \ks$. Therefore,
within the SM, $\beta$ as extracted from the CP asymmetry in $\phi \ks$
equals that as found in $\Psi \ks$ \cite{London:LonPec}. In fact, this is
true even if there are new-physics contributions to $\bd$-$\bdbar$ mixing.

However, since the $b\to s$ penguin is a pure loop effect, there can in
principle be significant new contributions from new physics
\cite{London:GrossWor}. Examples of such new physics include four
generations, non-minimal supersymmetry, and models with enhanced
chromomagnetic dipole operators. If there is new physics, the phase of the
decay amplitude may be changed. In this case, one will find $\beta$ (from
$\phi \ks$) is not equal to $\beta$ (from $\Psi \ks$). Therefore, by
measuring $\beta$ in $B_d(t) \to \phi \ks$, it might in fact be possible to
find $\alpha + \beta + \gamma \ne \pi$, even at $B$-factories. Note also
that, in addition to $\phi \ks$, the final states $\eta' \ks$, $\rho \ks$,
$\pi^0 \ks$, $\eta \ks$, {\it etc.} may be used. In fact, recent results
from CLEO, which show that the branching ratio for $B \to \eta' K$ is
larger than expected, indicate that this method of measuring $\beta$ may be
promising \cite{London:LonSoni}. 

The interesting thing is that what is really being probed here is new
physics in the $b\to s$ flavour-changing neutral current. This same new
physics will, in general, contribute to $\bs$-$\bsbar$ mixing. Thus, this
is in some sense a way of detecting new physics in $B_s$ mixing without
using $B_s$'s at all!

Having discussed this special case, I now return to the more conventional
ways of measuring the CP angles, via tree-level decays of $B$ mesons.
Suppose that the CP asymmetries are measured, and evidence for new physics
is found. What could this new physics be? We know that the new physics
contributes to $B^0$-${\overline{B^0}}$ mixing. Therefore a first step is
to classify models of new physics according to (i) whether they contribute
to $B^0$-${\overline{B^0}}$ mixing and (ii) if so, if new phases are
involved. 

Here is a fairly extensive list of models of new physics, along with a
discussion of their effects in $B^0$-${\overline{B^0}}$ mixing
\cite{London:GL}:
\begin{itemize}

\item Four generations: there are new loop-level contributions to the
mixing involving internal $t'$ quarks. Since the CKM matrix is now $4
\times 4$, new phases can be introduced.

\item $Z$-mediated flavour-changing neutral currents (FCNC's): if the
down-type quarks mix with an exotic vector singlet charge $-1/3$ quark,
then the flavour-changing couplings $Z b {\bar d}$ and $Z b {\bar s}$ will
be induced. In such models, there will be new contributions, with new
phases, to $B$ mixing through tree-level $Z$ exchange.

\item Multi-Higgs-doublet models:

\begin{itemize}

\item with natural flavour conservation (NFC): in such models there are new
contributions to $B$ mixing involving box diagrams with internal charged
Higgses. The charged Higgses couple to quarks through the CKM matrix, so no
new phases are introduced.

\item without NFC: in this case there can be tree-level FCNC's involving
the exchange of a neutral Higgs. Thus there are new contributions, with new
phases, in $B$ mixing.

\end{itemize}

\item Left-right symmetric models: except in the most fine-tuned models,
which I don't consider here, the mass of the $W_R$ is at least 1 TeV. This
renders its effects in $B$ mixing negligible. 

\item Supersymmetry:

\begin{itemize}

\item Minimal SUSY: there are many new contributions to $B$ mixing
involving box diagrams with internal supersymmetric particles. In the
minimal model, all couplings involve the CKM matrix, so that no new phases
are introduced.

\item Non-minimal SUSY: in non-minimal models, the new contributions can
also have new phases.

\end{itemize}

\end{itemize}

The above list shows that there are indeed many models of physics beyond
the SM which can contribute to $B^0$-${\overline{B^0}}$ mixing, some with
new phases, some without. The presence of such new physics will be detected
through measurements of the CP asymmetries. However, such measurements will
only tell us that new physics is present. While that would be a very
exciting development, we still would want to know what the new physics is.
How can we distinguish among the various possibilities listed above?

Some progress can be made through a simple observation. Any new physics
which affects $B^0$-${\overline{B^0}}$ mixing, which is a FCNC process, will
also in general affect the rare, flavour-changing decays $b\to s X$ and $b
\to d X$ (penguin decays). Therefore, by also looking at penguin decays, it
may be possible to identify some models of new physics \cite{London:GL}. In
fact, for certain types of new physics, if no deviation from the SM is
observed in penguin decays, this would rule out there being any effects in
$B$ mixing.

To see how this works, I will examine in detail one model of new physics:
$Z$-mediated FCNC's. As mentioned earlier, there are flavour-changing
couplings $Z b {\bar d}$ and $Z b {\bar s}$, parametrized by $U_{db}$ and
$U_{sb}$, respectively. These couplings are constrained by $BR(B \to
\mu^+\mu^- X) < 5 \times 10^{-5}$, leading to $|U_{qb}/V_{cb}| < 0.044$, or
$|U_{qb}| < 0.0017$. The new couplings $U_{qb}$ can have arbitrary phases.

In this model there are new, tree-level contributions to
$B^0$-${\overline{B^0}}$ mixing through $Z$ exchange. Comparing to the SM,
we find
\beq
{\Delta M_d^{\sss Z} \over \Delta M_d^{\sss W}} = (0.9~{\hbox{--}}~26)
\left[ {|U_{db}/V_{cb}| \over 0.04} \right]^2 ~,~~~~
{\Delta M_s^{\sss Z} \over \Delta M_s^{\sss W}} = 0.15 \left[
{|U_{sb}/V_{cb}| \over 0.04} \right]^2 ~. 
\eeq
Therefore $\bd$-$\bdbar$ mixing can be dominated by $Z$-FCNC's, with new
phases; $\bs$-$\bsbar$ mixing is still due mainly to $W$ box diagrams, but
the new contribution is non-negligible, so the new phases may be important.

Let us now examine the contribution of $Z$-FCNC's to penguin decays. The
constraint $|U_{qb}| < 0.0017$ is derived from the experimental limit $BR(B
\to \mu^+\mu^- X) < 5 \times 10^{-5}$. Therefore if the new coupling takes
its maximum allowed value, the model ``predicts'' the same branching ratio.
This is, in fact, a huge effect -- it is a smoking-gun signal. For the
$b\to s$ decay, this is roughly 10 times bigger than in the SM, while for
the $b\to d$ decay, it is an enhancement of about a factor of 100.
Furthermore, if the branching ratios for the decays $B \to X_s \ell^+
\ell^-$ and $B \to X_d \ell^+ \ell^-$ are found to be consistent with the
SM, this puts such stringent constraints on the $|U_{qb}|$ that it rules
out the possibility of any effects in $B^0$-${\overline{B^0}}$ mixing.

There are smoking-gun enhancements in other decays as well. For the
presently-allowed values of the $|U_{qb}|$,
\begin{itemize}

\item $BR(\bs \to \ell^+\ell^-)$ is enhanced by about a factor of 20.

\item $BR(\bd \to \ell^+\ell^-)$ is enhanced by about a factor of 300-400.

\item $BR(b \to s~{\rm EWP's})$ is enhanced by about a factor of 25.

\item $BR(b \to d~{\rm EWP's})$ is enhanced by about a factor of 500.

\end{itemize}
`EWP' stands for electroweak penguin decays. These are penguin decays which
are dominated in the SM by a virtual $Z$, {\it e.g.} $B^+ \to \phi \pi^+$
and $\bs \to \phi \pi^0$. There are no large effects of $Z$-mediated FCNC's
in $b\to s\gamma$ or in other hadronic penguins.

The point of all this is to demonstrate that, if there are significant
new-physics effects in $B^0$-${\overline{B^0}}$ mixing, then this same new
physics is also likely to have important effects in $B$ penguin decays.
Table 1 contains a summary of the effects of various models of new physics
on both $B^0$-${\overline{B^0}}$ mixing and penguin decays
\cite{London:GL}. 

\begin{table}[b!]
\caption{Contributions of models of new physics to $B^0$-${\overline{B^0}}$
mixing and $B$ penguin decays.}
\begin{tabular}{ldddd}
{\bf Model} & {\bf Contribution to} & {\bf New} & {\bf Contributions} &
{\bf Modes} \\
\ & $\bf B^0${\bf -}$\bf {\overline{B^0}}$ {\bf Mixing?} & {\bf Phases?} &
{\bf to Penguins?} & \ \\ 
\tableline
4 generations & Yes & Yes & Yes & EWP's \\
$Z$-FCNC's & Yes & Yes & Yes & $b\to q\ell^+ \ell^-$, \\
\ & \ & \ & \ & $B^0 \to \ell^+ \ell^-$, \\
\ & \ & \ & \ & EWP's \\
MHDM w/ NFC & Yes & No & Yes & $b\to s\ell^+ \ell^-$, \\
\ & \ & \ & \ & $B^0 \to \ell^+ \ell^-$, \\
MHDM w/o NFC & Yes & Yes & No & --- \\
Left-Right Symm. & No & --- & No & --- \\
MSSM & Yes & No & No & --- \\
Non-min. SUSY & Yes & Yes & Yes & ? \\
\end{tabular}
\end{table}

To sum up, there are many signals of new physics in CP asymmetries:
$Asym(\bs\to\Psi\phi) \ne 0$; $\alpha + \beta + \gamma \ne \pi$;
$Asym(\bd\to\Psi\ks) \ne Asym(\bd\to\phi\ks)$ [$\beta$]; $Asym(\bs\to
D_s^\pm K^\mp) \ne Asym(B^\pm \to D K^\pm)$ [$\gamma$]; $\alpha + \beta +
\gamma = \pi$ but $\alpha$, $\beta$ and $\gamma$ are inconsistent with the
SM (e.g. $\sin 2\beta < 0$); $\alpha + \beta + \gamma = \pi$, $\alpha$,
$\beta$ and $\gamma$ are consistent with the SM, but are inconsistent with
measurements of the sides of the UT; etc.

The main way in which new physics can enter is via new contributions to
$B^0$-${\overline{B^0}}$ mixing. (There is an exception: for pure penguin
decays, such as $\bd\to\phi\ks$, there can be new decay amplitudes.) There
are many models of new physics which can yield such new contributions. In
this talk I have considered four generations, $Z$-mediated FCNC's,
multi-Higgs-doublet models with and without natural flavour conservation,
minimal and non-minimal supersymmetry. 

Assuming that some signal for new physics is seen in the measurements of CP
asymmetries, one can partially distinguish among the various models by
looking at the rates for rare penguin decays. CP asymmetries and penguin
decays thus give complementary information regarding the identity of the
new physics.

\bigskip
\centerline{\bf Acknowledgements}
\bigskip

I would like to thank Dan Kaplan for the invitation to this excellent
conference. Thanks also to A. Ali, M. Gronau and A. Soni for pleasant
collaborations on some of the subjects discussed here, and to S. Sharpe for
correspondence regarding the appropriate value of $B_{\scriptscriptstyle
K}$. This work was financially supported by NSERC of Canada and FCAR du
Qu\'ebec.


\begin{references}

\bibitem{London:AliLon} The figure has been obtained following the
procedure described in A. Ali and D. London, \zpc{65}{95}{431}, {\it Nucl.\
Phys.\ (Proc.\ Suppl.)} {\bf 54A}, 297 (1997), except that I have taken
$B_{\scriptscriptstyle K} = 0.9 \pm 0.15$ (thanks to S. Sharpe for
discussions).

\bibitem{London:CPReview} For reviews, see, for example, Y. Nir and H.R.
Quinn in \stone, p.\ 520; I. Dunietz, {\it ibid.}, p.\ 550; M. Gronau, {\it
Proceedings of Neutrino 94, XVI International Conference on Neutrino
Physics and Astrophysics}, Eilat, Israel, May 29 -- June 3, 1994,  eds. A.
Dar, G. Eilam and M. Gronau,  {\it Nucl.\ Phys.\ (Proc.\ Suppl.)} {\bf
B38}, 136 (1995).

\bibitem{London:GL} For a more complete discussion, see M. Gronau and D.
London, \prd{55}{97}{2845}, and references therein.

\bibitem{London:DibLN} C.O. Dib, D. London and Y. Nir, \ijmp{6}{91}{1253}.

\bibitem{London:NirSilv} Y. Nir and D. Silverman, \npb{345}{90}{301}.

\bibitem{London:LonPec} D. London and R. Peccei, \plb{223}{89}{257}.

\bibitem{London:penguins} B. Grinstein, \plb{229}{89}{280}; M. Gronau,
\prl{63}{89}{1451}, \plb{300}{93}{163}.

\bibitem{London:GrossWor} Y. Grossman and M.P. Worah, \plb{395}{97}{241}.

\bibitem{London:LonSoni} D. London and A. Soni, hep-ph/9704277, to appear
in {\it Phys.\ Lett.} {\bf B}.

\end{references}
\end{document}